\documentclass[onecolumn,showpacs,showkeys,preprintnumbers,amsmath,amssymb]{revtex4}
\usepackage{setspace}
\usepackage{graphicx}
\usepackage{dcolumn}
\usepackage{bm}
\usepackage{epsfig}
\usepackage{subfigure}
\usepackage{subfigure}
\usepackage{subfigmat}

\begin{document}

\preprint{JPD:AP-IST 2009-Pinheiro}

\title[]{Electrical and Kinetic Model of an Atmospheric RF Device for Plasma
Aerodynamics Applications}

\author{Mario J. Pinheiro}\email{mpinheiro@ist.utl.pt}
\address{Department of Physics \& Institute for Plasma and Nuclear Fusion,
Instituto Superior Tecnico, Av. Rovisco Pais, 1049-001 Lisboa,
Portugal}

\author{Alexandre A. Martins}\email{aam@mail.ist.utl.pt}
\address{Institute for Plasma and Nuclear Fusion \&
Instituto Superior Tecnico, Av. Rovisco Pais, 1049-001 Lisboa,
Portugal}

\pacs{47.65.-d, 52.80.Pi, 51.50.Bv, 07.05.Tp, 47.85.LA}
\keywords{Magnetohydrodynamics and electrohydrodynamics, High-frequency and RF discharges, Electrical properties,
Computer modeling and simulation, Flow control}

\date{\today}

\begin{abstract}
The asymmetrically mounted flat plasma actuator is studied using a self-consistent 2-DIM
fluid model at atmospheric pressure. The computational model use the drift-diffusion approximation and a simple plasma phenomenological kinetic model. It is investigated its electrical
and kinetic properties, and calculated the charged species concentrations, surface charge density, electrohydrodynamic forces and gas speed. The present computational model contributes to understand the main physical mechanisms and methods to improve its performance.
\end{abstract}

\maketitle

\section{Introduction}

There has been a growing interest in the field of plasma
aerodynamics related to its outstanding importance in active flow
control, overriding the use of mechanical flaps
~\cite{Colver_96,Roth_03,Pinheiro_06,Moreau_07}. Plasma actuators create
a plasma above a blunt body that modify the laminar-turbulent
transition inside the boundary layer~\cite{Soldati_98,Soldati_99}, even at a high angle of attack~\cite{Roth_03,8}
they induce or reduce the fluid separation, and thus reducing
drag~\cite{Colver_96} and increasing lift~\cite{9,10,11}. They also allow sonic boom minimization
schemes~\cite{Sigelman_70,Boyd_99}, avoiding unwanted vibrations
or noise~\cite{12,13}, sterilizing or decontaminating surfaces and even frost removal~\cite{14} on any airfoil, jet engine or wind turbine, through pure electromagnetic control. Its promising potential extend to flow control at hypersonic speeds~\cite{Shin_07,Utkin_07} (while still using a jet-reaction aircraft propeller).

The asymmetric dielectric barrier discharge (DBD) plasma actuator is a normal glow discharge that, like all normal glow discharges, operates at the Stoletow point. This guarantees that the generation of the ion-electron pairs at one atmosphere is done efficiently.  In air, the minimum energy cost is 81 electron volts per ion-electron pair formed in the plasma.  In plasma torches, this energy cost can be of the order of one keV$/$ion-electron pair; in arcs, it can range from 10 to 50 keV$/$ion-electron pair.

Particle and fluid simulations have been done ~\cite{1,2,4} for a plasma actuator in pure oxygen and pure nitrogen showing the formation of an asymmetrical force that accelerates the ions dragging the neutral fluid in the direction of the buried electrode. A net force arises because the plasma density, and consequently, momentum transfer are greater during the second half of the bias cycle, due partially to the ion density greater a factor of 10 times during the second half-cycle. Thus, in each cycle there is created a total unidirectional force towards the buried electrode that can create neutral fluid flow velocities on the order of 8 m$/$s.

Two dimensional fluid models of a DBD plasma actuator have been made~\cite{5,6} which calculate the total force on ions and neutral particles and show that the force generated is of the same nature as the electric wind in a corona discharge, with the difference that the force in the DBD is localized in the cathode sheath region of the discharge and expands along the dielectric surface. While the intensity of this force is much larger than the existing force of a dc corona discharge, it is active during less than a hundred nanoseconds for each discharge pulse and, consequently, the time averaged forces are of the same magnitude in both cases.

The use of voltage pulses in plasma actuators, by modulation of the high frequency excitation voltage carrier wave by a square wave, introduces mean and unsteady velocity components and thus the air momentum is composed of both time-mean and oscillatory components of momentum, the overall effect improving momentum efficiency~\cite{15}.

Actuators placed on the leading edge of an airfoil can control the
boundary layer separation, while if located at the trailing edge can
control lift~\cite{4}. Enloe {\it et al.}~\cite{Enloe_03,Enloe_06} have found experimentally that
the thrust $T$ and maximum induced speed $u_{max}$ are proportional
to the input power $P$, which depends nonlinearly on the voltage
drop $\Delta V$ across the dielectric $T \propto u_{max} \propto P
\propto \Delta V^{7/2}$.

\begin{figure}
  \includegraphics[width=3.0 in]{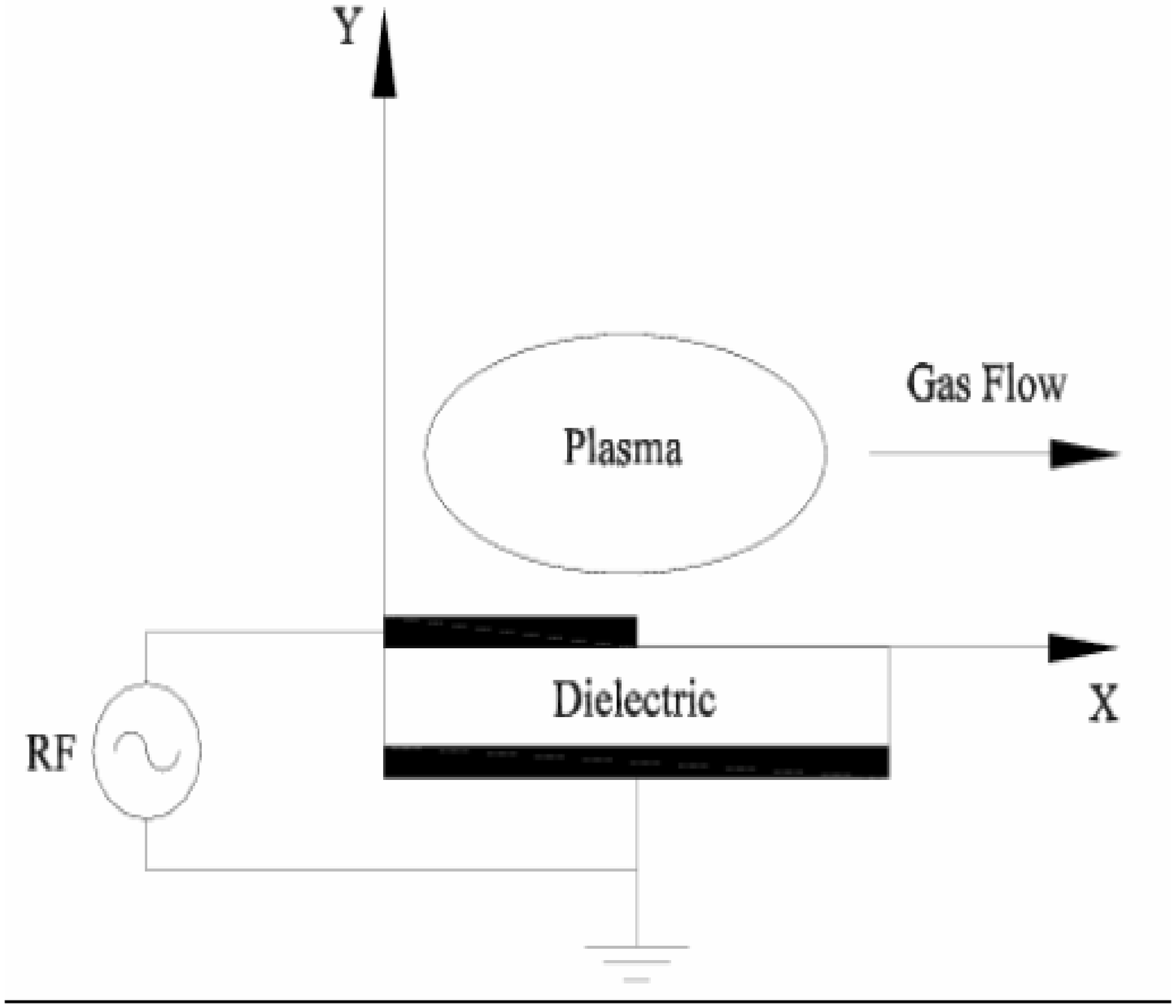}\\
  \caption{Schematic of the asymmetric plasma actuator.}\label{fig1}
\end{figure}

The aim of this paper is to present a self-consistent two dimensional modeling
of the temporal and spatial development of asymmetric DBD plasma actuator using
an EHD code (CODEHD) developed in our group. The computational model solves the governing equation in the drift-diffusion approximation and a plasma phenomenological kinetic model. Fig.~\ref{fig1} shows the rather simple configuration of the plasma actuator in coplanar configuration.

\section{Numerical model}
\subsection{Description}

At our knowledge the first comprehensive kinetic model of a dielectric barrier discharge plasma actuator was published by Singh {\it et al.} ~\cite{Singh}. Gadri~\cite{Gadri} has shown that an atmospheric glow
discharge is characterized by the same phenomenology as low-pressure
dc glow discharge. To subdue numerical complexity no detailed plasma chemistry with neutral heavy
species is presently addressed. At this stage it is only considered the kinetics
involving electrically charged species supposedly playing a
determinant role at atmospheric pressure: N$_2^+$, N$_4^+$, O$_2^+$,
O$_2^-$, and electrons. From the charged species populations and as well the
electric field controlling their dynamics, it can be studied electrohydrodynamics (EHD) with interest to plasma actuators, like the body forces acting on the plasma horizontally (neutral flow control) and perpendicularly (boundary-layer control) to the energized electrode; and the neutral particles average speed.

The applied voltage has a sinusoidal wave form $V(t)=V_{dc}+V_{rms} \sin(\omega t)/\sqrt{2}$, where the root mean square voltage, $V_{rms}$, in this case of study is 5 kV and the applied frequency is $f=5$ kHz. Therefore, the dynamical time is $T=200$ $\mu$s.

The plasma actuator simulation domain is a 2-Dim Cartesian geometry with the total length along the
Ox-axis $L_x=4$ mm and height $L_y=4$ mm. It consists of conductive copper
strips (with negligible thickness) of width $w=1$ mm, separated by a
0.065 cm thick dielectric with width equal to 3 mm and relative dielectric
permittivity $\epsilon_r=5$. The electrical capacity of the reactor is given by the
conventional formula $C=\epsilon_r \epsilon_o S/d$.

\subsection{Transport parameters and rate coefficients}

The working gas is a ``airlike" mixture of a
fixed fraction of nitrogen ([N$_2]/$N$=.78$) and oxygen ([O$_2]/$N$=0.22$), as
is normally present at sea level at $p=1 atm$. The electron
homogeneous Boltzmann equation is solved with the 2-term expansion
in spherical harmonics~\cite{Ferreira} for a mixture of N$_2-$O$_2$(22$\%$), assumed thoroughly constant.
The gas temperature is also assumed constant, both spatially and in time frame, with $T_g=300$ K, and the same applies to the
vibrational temperature of nitrogen $T_v (N_2)=2000$ K and oxygen
$T_v(O_2)=2000$ K. This assumption avoids the need of a more complex vibrational kinetic model.

Using the set of cross sections of excitation by electron impact taken from
Ref.~\cite{Siglo} rates coefficients and transport parameters needed for the electronic kinetics are obtained. So far, the species included in the
present model are the following: $N_4^+$, $N_2^+$, $O_2^+$, $O_2^-$ and electrons. Remark that at 1
atm the concentrations of $N_4^+$ ions need to be introduced since they are bigger than $N_2^+$, due
partially to the reaction $N_2^+ + N_2 \to N_4^+$, which occurs at a
higher rate than the direct ionization. Also, notice that at $p < 10^{-2}$ Torr loss by
ambipolar diffusion becomes overwhelming important. Ion diffusion and mobility coefficients were taken from the Report
\cite{Sigmond}: $\mu_{O_2^-} .N= 6.85 \times 10^{21}$ V$^{-1}$
m$^{-1}$ s$^{-1}$ (on the range of E/N with interest here),
$\mu_{O_2^+} N = 6.91 \times 10^{21}$ V$^{-1}$ m$^{-1}$ s$^{-1}$,
and $\mu_{N_2^+} N = 5.37 \times 10^{21}$ V$^{-1}$ m$^{-1}$s$^{-1}$. The gas density at $p=1$ atm and assuming $T_g=300$ K is $N=2.447 \times 10^{25}$ $\#/m^{3}$.

At atmospheric pressure the local equilibrium assumption holds and the transport coefficients ($\nu_{ion}^{N_2}$, $\nu_{ion}^{O_2}$,
$\mu_e$, $\mu_p$, $D_e$, $D_p$) depend on space and time
$(\textbf{r},t)$ only through the local value of the electric
field $\mathbf{E}(\mathbf{r},t)$; this is the so called
{\it hydrodynamic regime}.

To avoid the use of Navier-Stokes equations and to obtain a faster numerical solution of the present hydrodynamic
problem it is assumed that the gas flow does not alter the plasma
characteristics and is much smaller than the charged particle drift
velocity. This assumption allows a simplified description of the
flow.

With the above assumptions charged species can be described by
continuity equations and momentum transport equations in the
drift-diffusion approximation. This last approximation is valid if
their drift energy is negligible with respect to thermal energy.
Also, in the drift-diffusion equation it is neglected the
temperature gradient term (e.g., Ref.~\cite{6}).

\begin{table*}
\caption{\label{tab:table2} List of reactions taken into account with CODEHD. Rate coefficients for chemical processes were taken from Ref.~\cite{Kossyi92}.}
\begin{ruledtabular}
\begin{tabular}{ccc}
 Kind of reaction &  Process & Rate coefficient\\
\hline
Ionization & $e + N_2 \to 2e + N_2^+$ &  $\nu_{ion}^{N_2}$ \footnotemark[1]\\
Ionization & $e + O_2 \to 2e + O_2^+$ &  $\nu_{ion}^{O_2}$ \footnotemark[1]\\
3-body electron attachment & $e + O_2 + O_2 \to O_2^- + O_2$ & $K_{a1}=1.4 \times 10^{-29} (\frac{300}{T_e})\exp(-600/T_g) K_1(T_g,T_e)$  (cm$^6/$s) \footnotemark[2] \\
3-body electron attachment & $e + O_2 + N_2 \to O_2^- + N_2$ & $K_{a2}=1.07 \times 10^{-31} (\frac{300}{T_e})^2 K_2(T_g,T_e)$ (cm$^6/$s) \footnotemark[3]\\
Collisional detachment & $O_2^- + O_2 \to  e + 2 O_2$   & $K_{da}=2.9 \times 10^{-10} \sqrt{\frac{T_g}{300}} \exp(-5590/T_g)$ (cm$^3/$s)  \\
e-ion dissociative recombination & $ N_2^+ + e \to 2N$   &  $\beta=2.8 \times 10^{-7} \sqrt{\frac{300}{T_g}}$ (cm$^3/$s) \\
e-ion dissociative recombination & $O_2^+ + e \to 2O$    & $\beta=2.8 \times 10^{-7} \sqrt{\frac{300}{T_g}}$ (cm$^3/$s) \\
2-body ion-ion recombination & $O_2^- + N_2^+ \to O_2 + N_2$  & $\beta_{ii}=2 \times 10^{-7} \sqrt{\frac{300}{T_g}}[1+10^{-19} N (\frac{300}{T_g})^2]$ (cm$^3/$s) \\
2-body ion-ion recombination & $O_2^- + O_2^+ \to 2 O_2$      & $\beta_{ii}=2 \times 10^{-7} \sqrt{\frac{300}{T_g}}[1+10^{-19} N (\frac{300}{T_g})^2]$ (cm$^3/$s) \\
Ion-conversion & $N_2^+ + N_2 + N_2 \to N_4^+ + e$ & $K_{ic1}=5 \times 10^{-29}$ (cm$^6/$s) \\
Recombination & $N_4^+ + e \to 2 N_2$ & $K_{r2}=2.3 \times 10^{-6}/(Te/300)^{0.56}$ \\
Ion-conversion & $N_4^+ + N_2 \to N_2^+ + 2N_2$ &  $K_{ic2}=2.1\times 10^{-16} \exp(T_g/121)$ (cm$^3/$s) \\
\end{tabular}
\end{ruledtabular}
\footnotetext[1]{Numerical data obtained by solving the quasi-stationary,
homogeneous electron Boltzmann equation. See Ref.~\cite{Ferreira}
for details.}

\footnotetext[2]{With $K_1=\exp(700(T_e-T_g)/(T_eT_g))$.}
\footnotetext[3]{With $K_2=\exp(-70/T_g)\exp(1500(T_e-T_g)/(T_e
T_g))$.}
\end{table*}

The reactions included in the present ``effective" kinetic model are listed in
Table ~\ref{tab:table2}. It is assumed that all volume ionization
is due to electron-impact ionization from the ground state and the
kinetic set consists basically in ionization, attachment and
recombination processes. The kinetics of excited states and heavy
neutral species is not considered, particularly the possible important role of metastable species (certainly improving plasma density through Penning ionization~\cite{Massines_99}) and nitrous oxide (poisoning gas) are neglected. The major drawback associated to the present ``effective" kinetic model is the lack of consistency due to the neglect of atomic species, but it allows to investigate the general trends of the plasma actuator operating characteristics.

\begin{figure}
  \includegraphics[width=5.0 in]{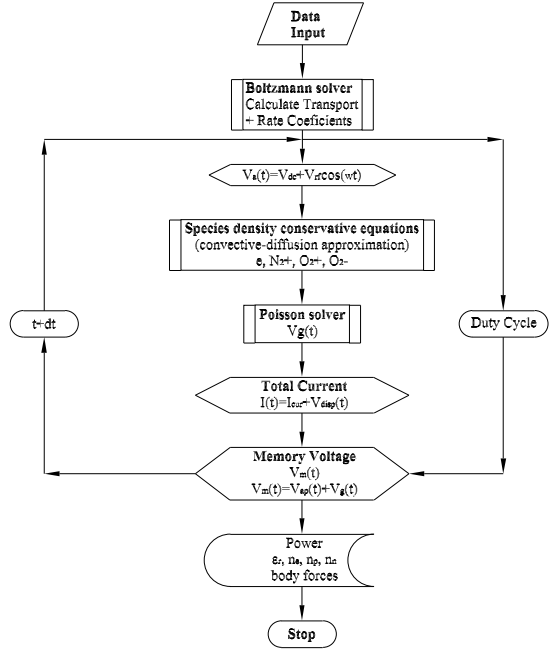}\\
  \caption{Program flow chart of the EHD code (CODEHD).}\label{orga}
\end{figure}

For more concise notation, we put $n_{p2}\equiv
[N_4^+]$; $n_{p1}\equiv [N_2^+]$; $n_p \equiv [O_2^+]$; $n_n
\equiv [O_2^-]$, and $n_e \equiv [e]$.

With the above assumptions charged species can be described by
continuity equations and momentum transport equations in the
drift-diffusion approximation. This last approximation is valid if
their drift energy is negligible with respect to thermal energy.
Also, in the drift-diffusion equation it is neglected the
temperature gradient term.

\subsection{Numerical model}

The particle's governing equations are of convection-diffusion type
and they are solved using a well-known finite-element method proposed by Patankar~\cite{Patankar}. Poisson equation for the electric field is solved applying the successive over-relaxation method (SOR) for boundary-value problem. The
chosen time step is limited by the value of the dielectric
relaxation time. For the present calculations it was used (100x100)
computational meshes. The entire set of equations are integrated
successively in time supposing the electric field constant during
each time step, obtaining a new value of the electric field after
the end of each time step. The method used to integrate the
continuity equations and Poisson equation was assured to be
numerically stable, constraining the time step width to the well
known Courant-Levy-Friedrich stability criterion.

\begin{figure}
  \includegraphics[width=3.5 in, height=3.0 in]{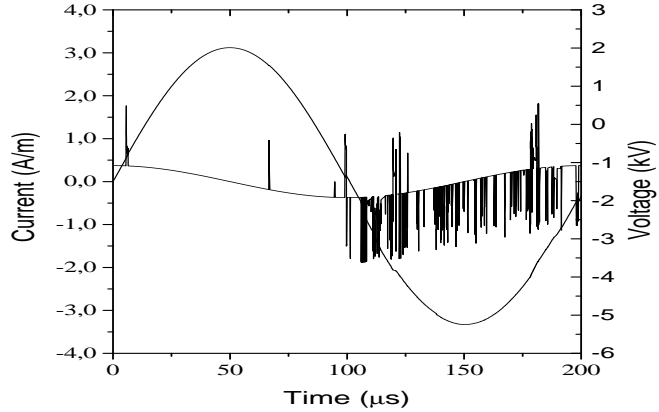}\\
  \caption{Electric current, applied voltage, gas voltage and memory voltage as a function of time.
  Conditions: Case I. Solid curve: current; dot curve: $V_m$; dashed-dot curve: $V_g$; dashed curve: $V$. Conditions: $V_{rms}$=5 kV, f=5 kHz,
dielectric width=3 mm. Solid curve: current; dot curve: $V_m$;
dashed-dot curve: $V_g$; dashed curve: V.}\label{fig3}
\end{figure}

The governing equation for N$_4^+$ (at atmospheric pressure the
nitrogen ion predominant is N$_4^+$) is:
\begin{equation}\label{Eq1}
\frac{\partial n_{p2}}{\partial t} + \mathbf{\nabla} \cdot (n_{p2}
\mathbf{v}_{p2}) = \delta_{N_2}^2 N^2 n_{p1} K_{ic1} - K_{ic2} \delta_{N_2} N
n_{p2}- K_{r2} n_{p2} n_e.
\end{equation}
The governing equation for N$_2^+$ (one of the most mobile charged specie in the plasma) is:
\begin{widetext}
\begin{equation}\label{Eq2}
\frac{\partial n_{p1}}{\partial t} + \mathbf{\nabla} \cdot (n_{p1}
\mathbf{v}_{p2}) =  n_e \nu_{ion}^{N_2} + K_{ic2} \delta_{N_2} N n_{p2} -
\beta_{ii} n_n n_{p1} - \beta n_e n_{p1} - K_{ic1} \delta_{N_2}^2 N^2 n_{p1}.
\end{equation}
\end{widetext}
The oxygen positive ion considered is O$_2^+$ and its resultant governing
equation is given by:
\begin{equation}\label{Eq3}
\frac{\partial n_{p}}{\partial t} + \mathbf{\nabla} \cdot (n_{p}
\mathbf{v}_{p}) =  n_e \nu_{ion}^{O_2} - \beta_{ii} n_n n_p -
\beta n_e n_p.
\end{equation}
The negative ion O$_2^-$ was
introduced (due to high electronegativity of oxygen), and its governing equation was written as:
\begin{equation}\label{Eq4}
\frac{\partial n_{n}}{\partial t} + \mathbf{\nabla} \cdot (n_{n}
\mathbf{v}_{n}) = \nu_{att}^{O_2}n_e - \beta_{ii} n_{p1} n_n - K_{da}
\delta_{O_2} N n_n.
\end{equation}
Finally, the governing equation for electrons can be written in the
form:
\begin{widetext}
\begin{equation}\label{Eq5}
\frac{\partial n_{e}}{\partial t} + \mathbf{\nabla} \cdot (n_{e}
\mathbf{v}_{e}) = n_e (\nu_{ion}^{N_2} + \nu_{ion}^{O_2} -
\nu_{att}^{O_2}) - \beta n_e (n_p + n_{p1}) + K_{da} \delta_{O_2} N n_n -
K_{r2} n_{p1} n_e.
\end{equation}
\end{widetext}
Photo-ionization was not included as a non-local secondary effect.

In order to avoid the use of Navier-Stokes equation, the {\it drift-diffusion
approximation} for the charged particle mean velocities appearing
in the continuity equations is used instead:
\begin{equation}\label{Eq6}
n_i \mathbf{v}_i = n_i \mu_i \mathbf{E} - \nabla (n_i D_i).
\end{equation}
Here, $\mu_i$ and $D_i$ represent the charged particle mobility
and the respective diffusion coefficient. The applied voltage has
a sinusoidal wave form:
\begin{equation}\label{Eq7}
V(t) = V_{dc} + V_0 \sin (\omega t),
\end{equation}
where $V_{dc}$ is the dc bias voltage (although here we fixed to
ground, $V_{dc}=0$) and $\omega$ is the applied angular frequency.
$V_0$ is the maximum amplitude with the root mean square voltage
in this case of study $V_{rms}=5$ kV and the applied frequency
$f=5$ kHz.

In an RF discharge the calculation of the displacement current is
a not a simple matter. However, the total current (convective plus
displacement current) was determined using the following equation given by Sato and Murray
\cite{Sato}:
\begin{widetext}
\begin{equation}\label{Eq8}
I_d(t)=\frac{e}{V} \int_V \left( n_p \mathbf{w}_p - n_e
\mathbf{w}_e - n_n\mathbf{w}_n - D_p \frac{\partial n_p}{\partial
z} + D_e \frac{\partial n_e}{\partial z} + D_n \frac{\partial
n_n}{\partial z} \right) \cdot \mathbf{E}_L d v +
\frac{\epsilon_0}{V} \int_V \left( \frac{\partial
\mathbf{E}_L}{\partial t}\cdot \mathbf{E}_L \right) dv,
\end{equation}
\end{widetext}
where $\int_V dv$ is the volume occupied by the discharge,
$\mathbf{E}_L$ is the space-charge free component of the electric
field. The last integral when applied to our geometry gives the
displacement current component
\begin{equation}\label{Eq9}
I_{disp} (t) = \frac{\varepsilon_0}{d^2} \frac{\partial
V}{\partial t} \int_V dv.
\end{equation}

\begin{figure}
  \includegraphics[width=3.0 in]{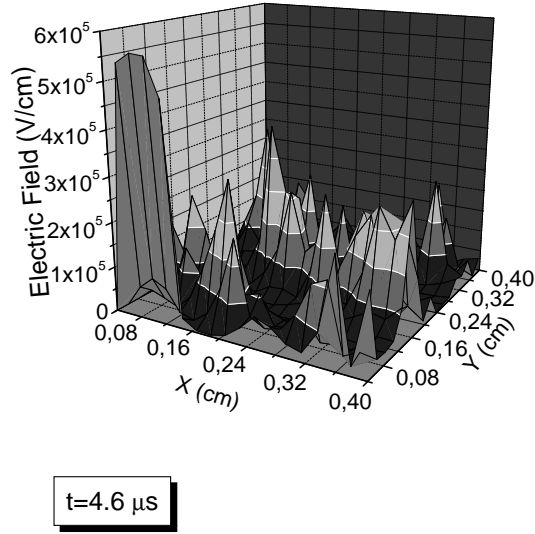}\\
  \caption{Electric field distribution in the X-Y plane at a given instant of time during the first-half
  cycle.}\label{f6}
\end{figure}

We assume throughout the calculations $\gamma=5 \times 10^{-2}$ since Auger electrons are assumed to be produced by impact of positive
ions on the cathode with such an efficiency. Hence, the flux density of secondary electrons out of the cathode
is given by
\begin{equation}\label{Eq10}
\mathbf{j}_{se} (t) = \gamma \mathbf{j}_p (t),
\end{equation}
with $\mathbf{j}_p$ denoting the flux density of positive ions. Secondary electron emission plays a fundamental role on the working
of the asymmetric DBD plasma actuator. The progressive accumulation of electric charges over
the dielectric surface develops a so-called ``memory voltage", whose expression is given by:
\begin{equation}\label{Eq11}
V_m(t)=\frac{1}{C_{ds}} \int_{t_0}^t I_d (t') d t' + V_m(t_0).
\end{equation}
Here, $C_{ds}$ is the equivalent capacitance of the discharge.

As charged particles are produced in the plasma volume, it is calculated the space-charge electric field by solving the
Poisson equation coupled the particle's governing equations:
\begin{equation}\label{Eq12}
\Delta V = -\frac{e}{\epsilon_0} (n_p - n_e - n_n).
\end{equation}
We assumed the following boundary conditions:
\begin{itemize}
    \item over the electrode (Dirichlet boundary condition): $V(x,y=0,t)=V-V_m$;
    \item over the insulator (Neumann boundary condition): $E_n=(\mathbf{E} \cdot
    \mathbf{n})=\frac{\sigma}{2 \epsilon_0}$.
\end{itemize}
The flux of electric charges impinging on the dielectric surface
builds up a surface charge density $\sigma$ which was calculated by
balancing the flux to the dielectric and it is governed by
\begin{equation}\label{Eq13}
\frac{\partial \sigma}{\partial t}=e
(|\Gamma_{p,n}|-|\Gamma_{e,n}|).
\end{equation}
Here, $\mathbf{\Gamma}_{p,n}$ and $\mathbf{\Gamma}_{e,n}$
represent the normal component of the flux of positive and
negative ions and electrons to the dielectric surface.
Furthermore, it is assumed that ions and electrons recombine
instantaneously on the perfectly absorbing surface. As it will be discussed later, this simplified assumption constitute a drawback of the present model.

The entire set of equations were solved together self-consistently
at each 1 ns time step, as illustrated with the program flow chart shown in Fig.~\ref{orga}.

\section{Results}

The electrohydrodynamic area of research has grown to a large extent lately, but it still remains to achieve a better understanding on how the
charged particles transfer momentum to neutrals, and what physical
limitations restrain the applicability of this device for boundary
control and neutral flow propulsion.

The simulations were done for a two-dimensional flat staggered
geometry, while assuming the plasma homogeneous along the OZ-axis (see Fig.~\ref{fig1}).
This is essentially a ``surface discharge" arrangement with
asymmetric electrodes.

\subsection{Electrical characteristics}

Fig.~\ref{fig3} shows the evolution along a full period of the
calculated electric current (convective plus displacement currents), applied voltage, gas voltage and memory
voltage. A stationary solution of the entire set of
equations modeling the apparatus is achieved typically after 4-5 cycles with CODEHD. With
the assumed conditions at about 700 Volts (or $V_g=2023$ Volts) electron avalanches
develop (the first avalanches occur in the gas gap), replenishing the volume above the surface with charged
particles. In a coplanar discharge, the current density increases with the applied voltage and almost nearly with the permittivity~\cite{Gibalov_04}.

This corresponds to roughly $E_{ion}=V_g/d=31$ kV$/$cm, when the theoretical dielectric breakdown of air is 30 kV$/$cm. Hence, the charged particles flowing to the dielectric
surface start accumulating on the surface, building-up an electric
field that prevents the occurrence of a high current, and therefore
quenching the discharge development at an early stage. At about 3 $\mu$s develops the first avalanche. The pulse form is typical: the first peak corresponds to the arrival of fast electrons to the electrode and it corresponds to the instant when the electric field is maximal; afterwards, the pulse slowly fade out, due to the dielectric charging and consequent decrease of the electric field.

Also, from Fig.~\ref{f6} it is clear that the electric field has a maximum
peak at the edge of the exposed (and energized) electrode. From this
edge the successive streamers (current pulses) are initiated.

At 4$th$ cycle the transient regime is totally damped and the
electrical characteristics stabilize. During the first (positive)
half-cycle electrons and negative ions are formed above the dielectric surface, which
acts as a virtual cathode (buried below the dielectric), while
they are pushed by the electric field towards the positive
(energized) electrode located on the top of the dielectric playing the role
of the anode, as can be seen from the set of snapshots shown from
Figs.~\ref{f:small_multiple} and~\ref{f:small_multiple2}. The dielectric surface is progressively charged.

\begin{figure}
  \includegraphics[width=3.0 in]{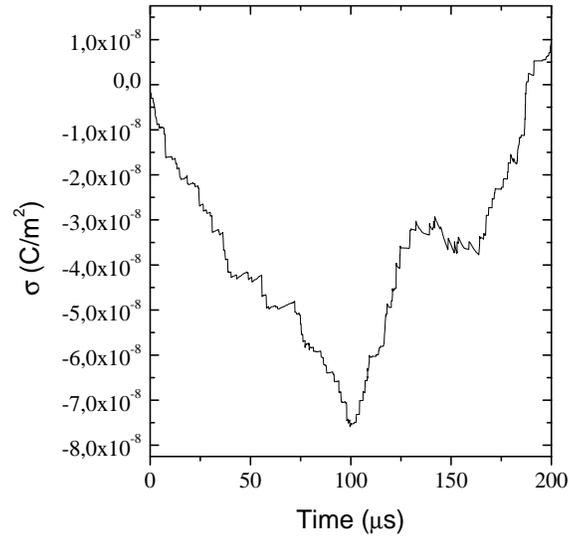}\\
  \caption{Surface charge density as a function of time.
  Conditions: Vrms=5 kV, f=5 kHz, dielectric width=3 mm.}\label{fig3a}
\end{figure}

Fig.~\ref{fig3a} shows the surface charge density vs. time, for the case of a larger dielectric width. The charge density should be of the order $\sigma \simeq 10^{-8}$ C$/$m$^2$, since $Q=CV=\sigma S$, or $\sigma=CV/S=\epsilon_o\epsilon_rV/d$. It becomes increasingly negative during the first half-cycle and tends to reverse sign during the second half-cycle. The memory voltage is regulated by the surface charge density deposited over the dielectric by the successive streamers. Notice that when the energized electrode is negative, more streamers occur.

\begin{figure}
 \begin{subfigmatrix}{2}
  \subfigure{\includegraphics[width=2.75 in, height=3.75 in]{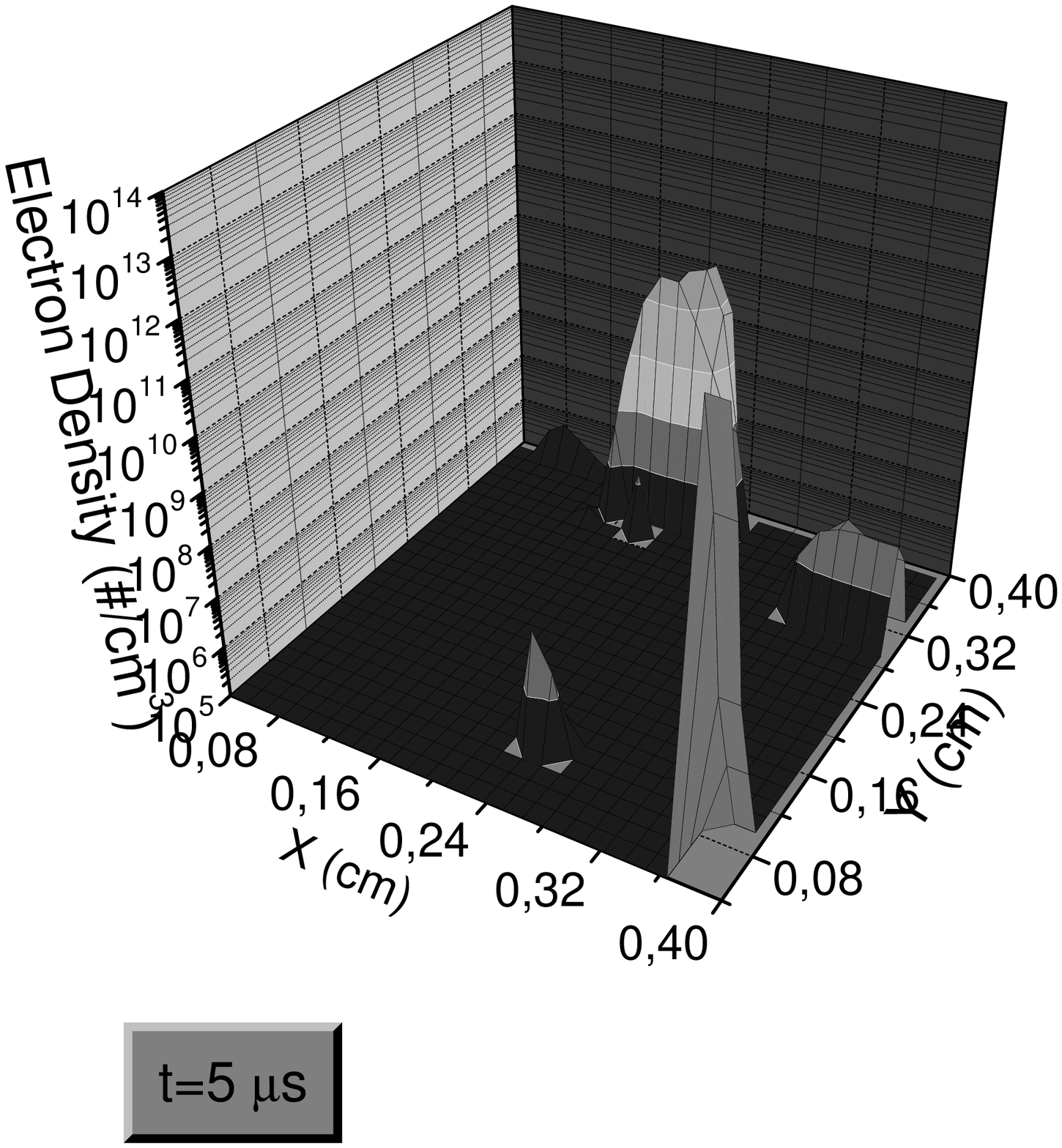}}
  \subfigure{\includegraphics[width=2.75 in, height=3.75 in]{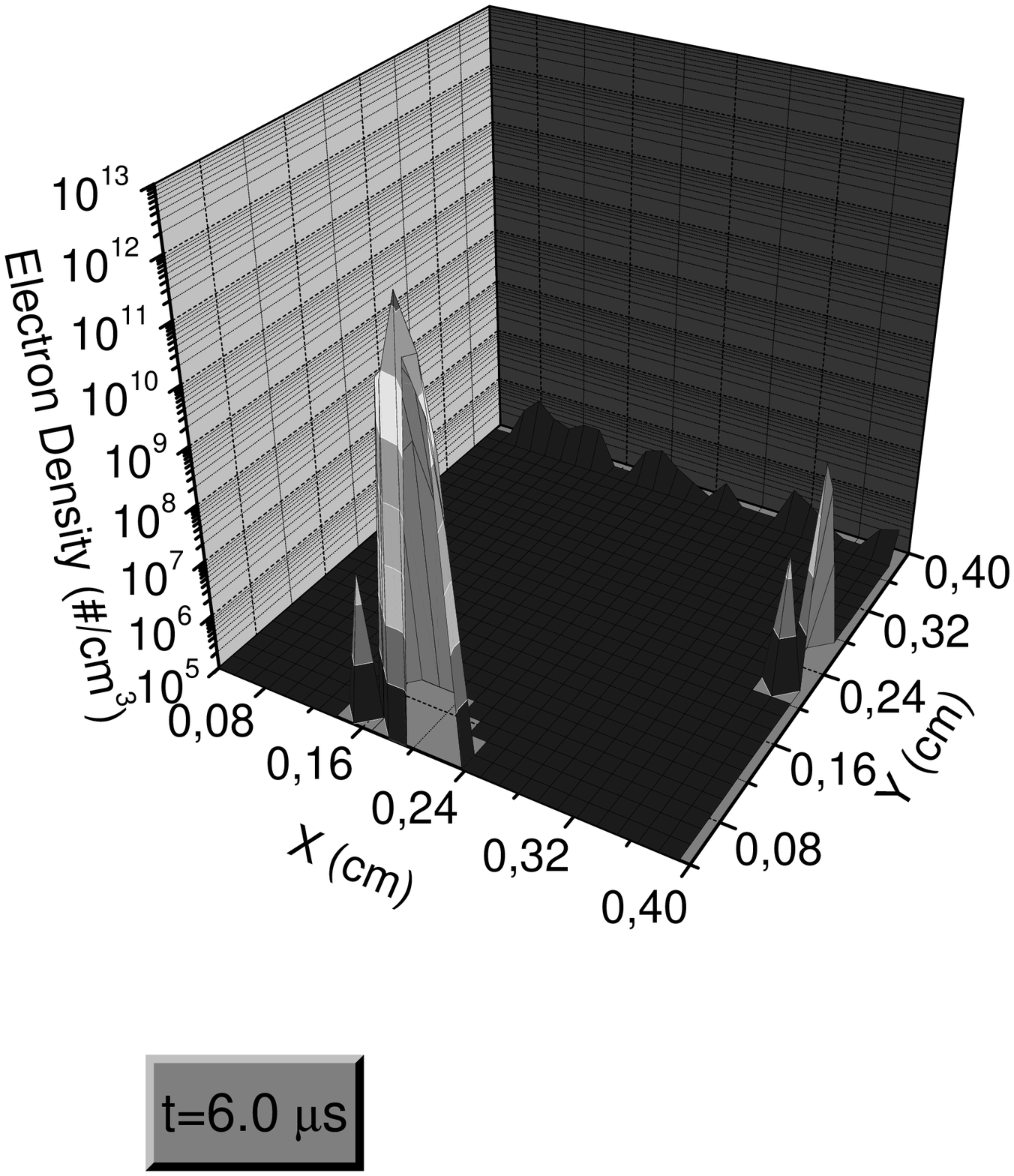}}
 \end{subfigmatrix}
 \caption{A time series shown of the electrons density at the
 threshold of an avalanche during the first half-cycle.}
 \label{f:small_multiple}
\end{figure}

\begin{figure}
 \begin{subfigmatrix}{2}
  \subfigure{\includegraphics[width=2.75 in, height=3.75 in]{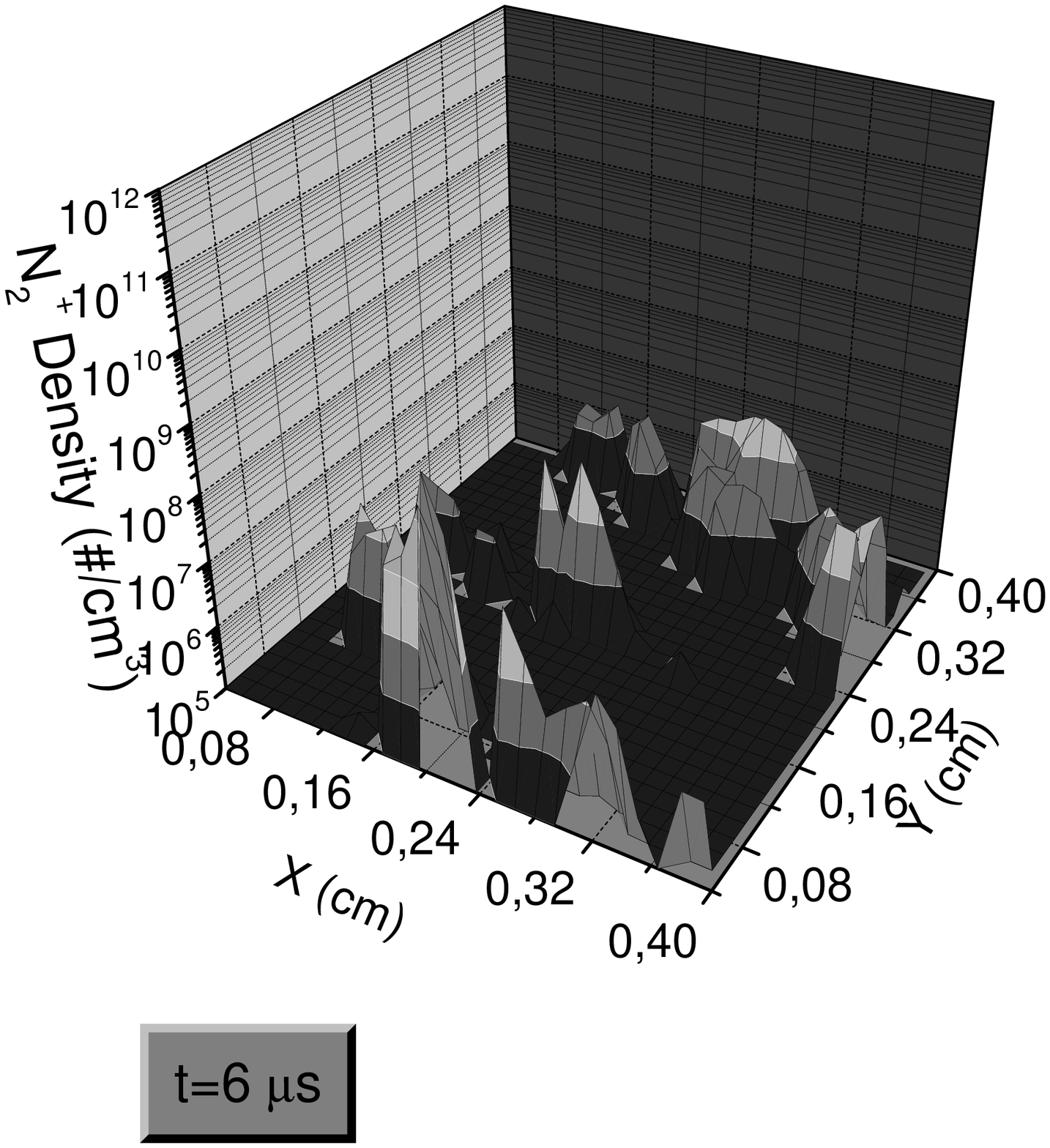}}
  \subfigure{\includegraphics[width=2.75 in, height=3.75 in]{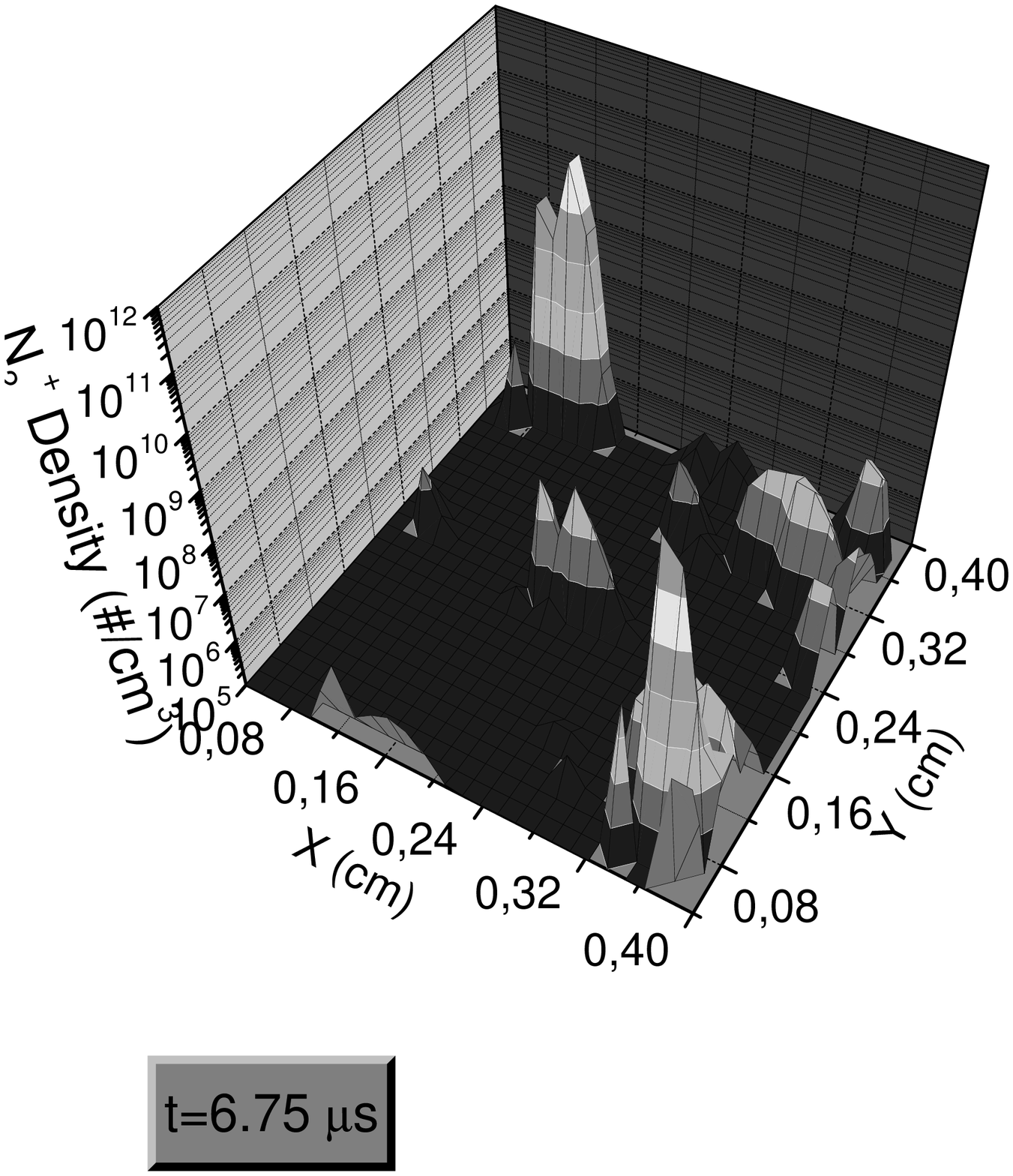}}
 \end{subfigmatrix}
 \caption{Two dimensional distribution of $N_2^+$ ions particle density predicted by
 the numerical code for typical conditions, during an avalanche occurring at the first half-cycle.}
 \label{f:small_multiple2}
\end{figure}

Fig.~\ref{f:small_multiple} shows a snapshot of the electrons
density along time since the beginning of an electron streamer
occurring during the first-half cycle. The density profiles were
obtained after 4-5 cycles with the dielectric charging onset
resulting in the formation of a virtual negative electrode.
Electrons are driven along the electric field lines toward the anode
(the energized electrode at the first-half cycle).

\begin{figure}
  \includegraphics[width=3.5 in]{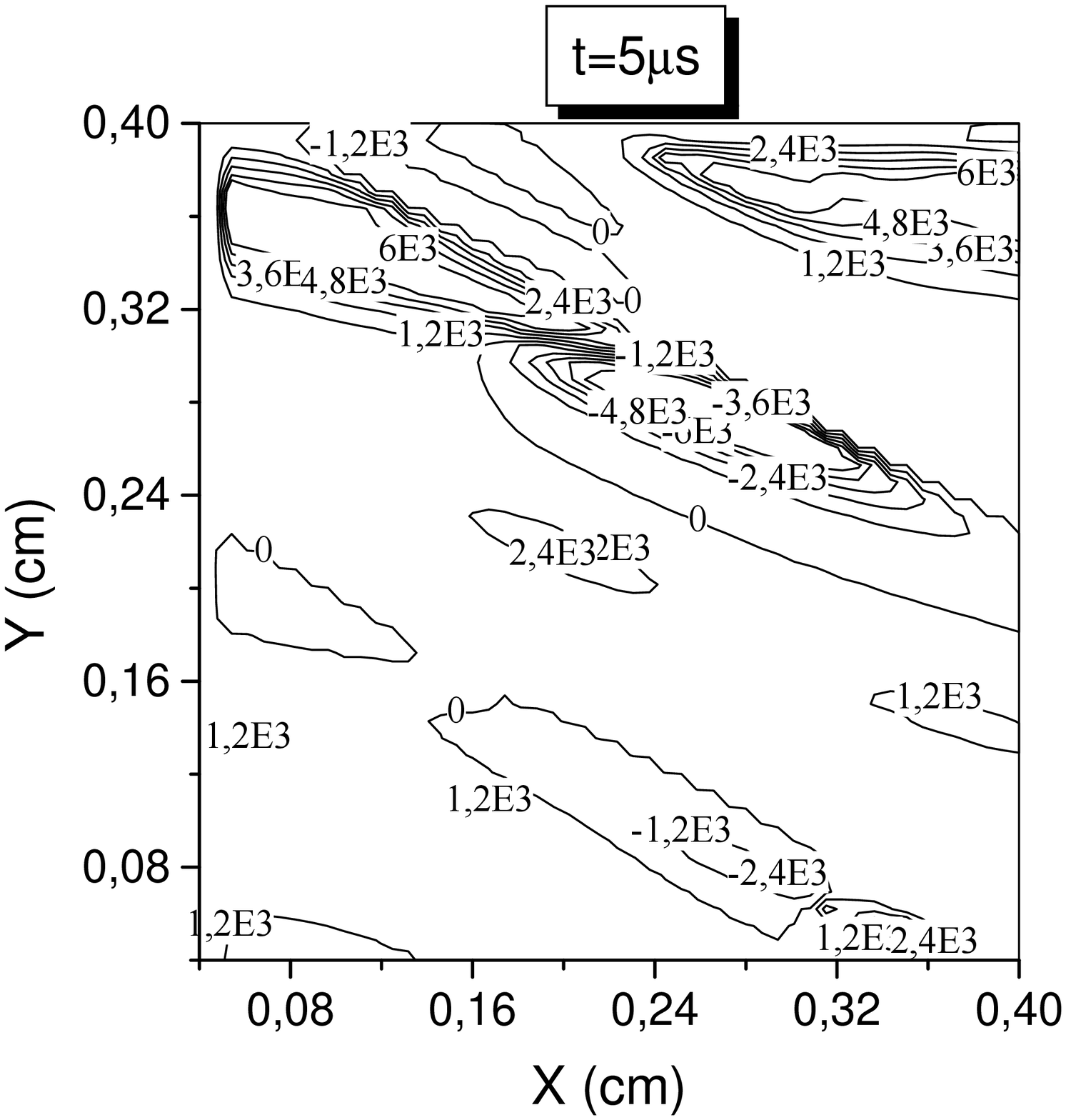}\\
  \caption{Contours of constant potential at $t=5 \mu$s. Same conditions as in
  Fig.3. Compare with Figs.4-5.}\label{fig30}
\end{figure}

At the same time and in the reversed direction a streamer of
positive ions $N_2^+$ are driven to the cathode, as it is shown at
the time series illustrated in Fig.~\ref{f:small_multiple2}. During
the streamers ponderomotive forces attain their highest magnitude. The event occurring at the instant of time
$t=6.75 \mu$s portrays the arrival of ions N$_2^+$ to the virtual cathode.

\begin{figure}
  \includegraphics[width=3.0 in]{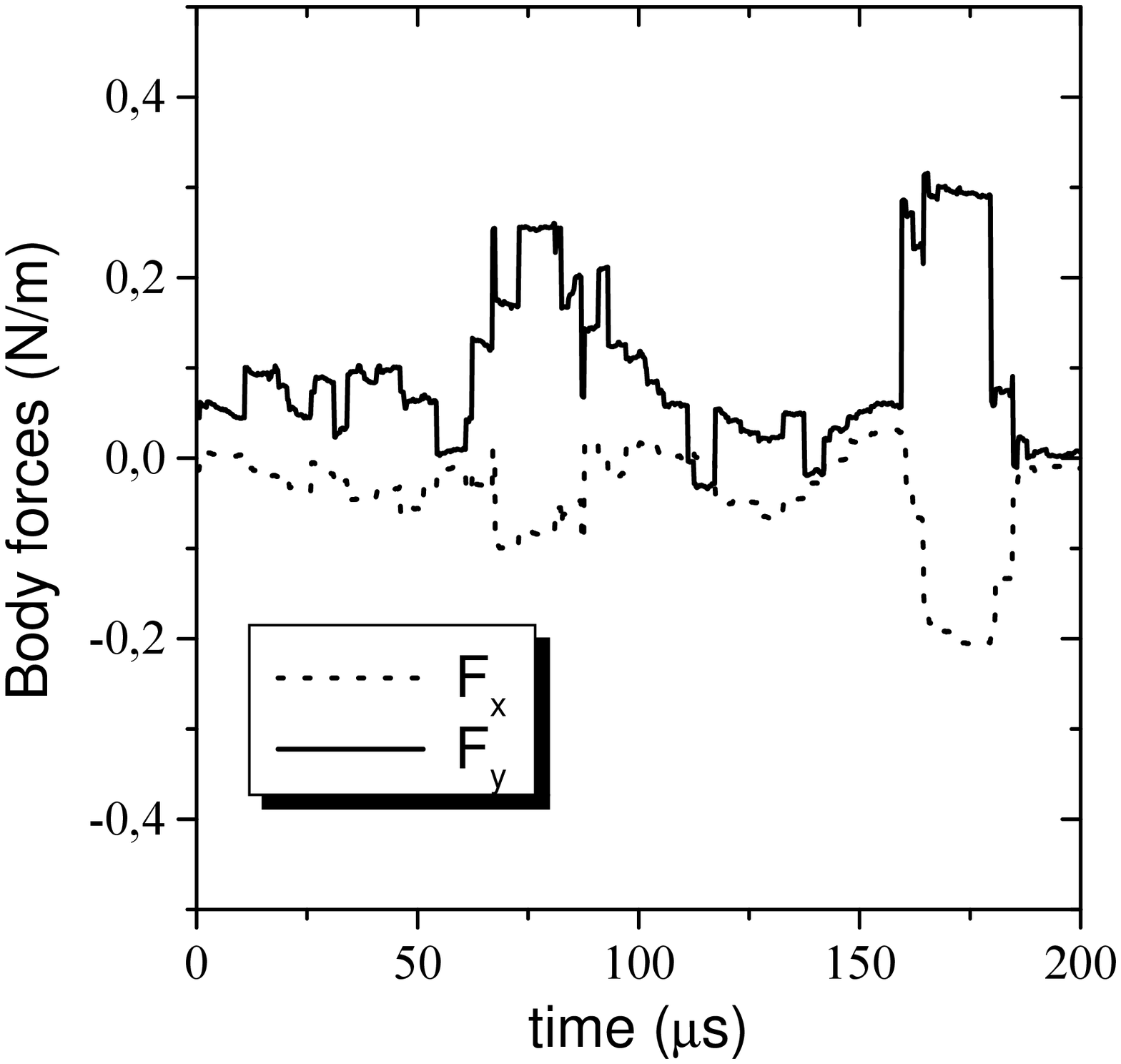}\\
  \caption{Electrohydrodynamic forces (in N$/$m) as a function of time.
  Same conditions as in Fig.~\ref{fig3}.}\label{fig4}
\end{figure}

We notice that the $N_2^+$ ions are mainly created in volume and are
immediately driven to the virtual cathode, while an average number
of them are retained nearby the anode decreasing almost linearly
with distance from the energized electrode surface. Thus, it is
clear that the propagation of the electron avalanche near (and
above) the dielectric surface is of considerable importance,
dictating the strength of the body forces and gas speed.

\begin{figure}
  \includegraphics[width=3.0 in]{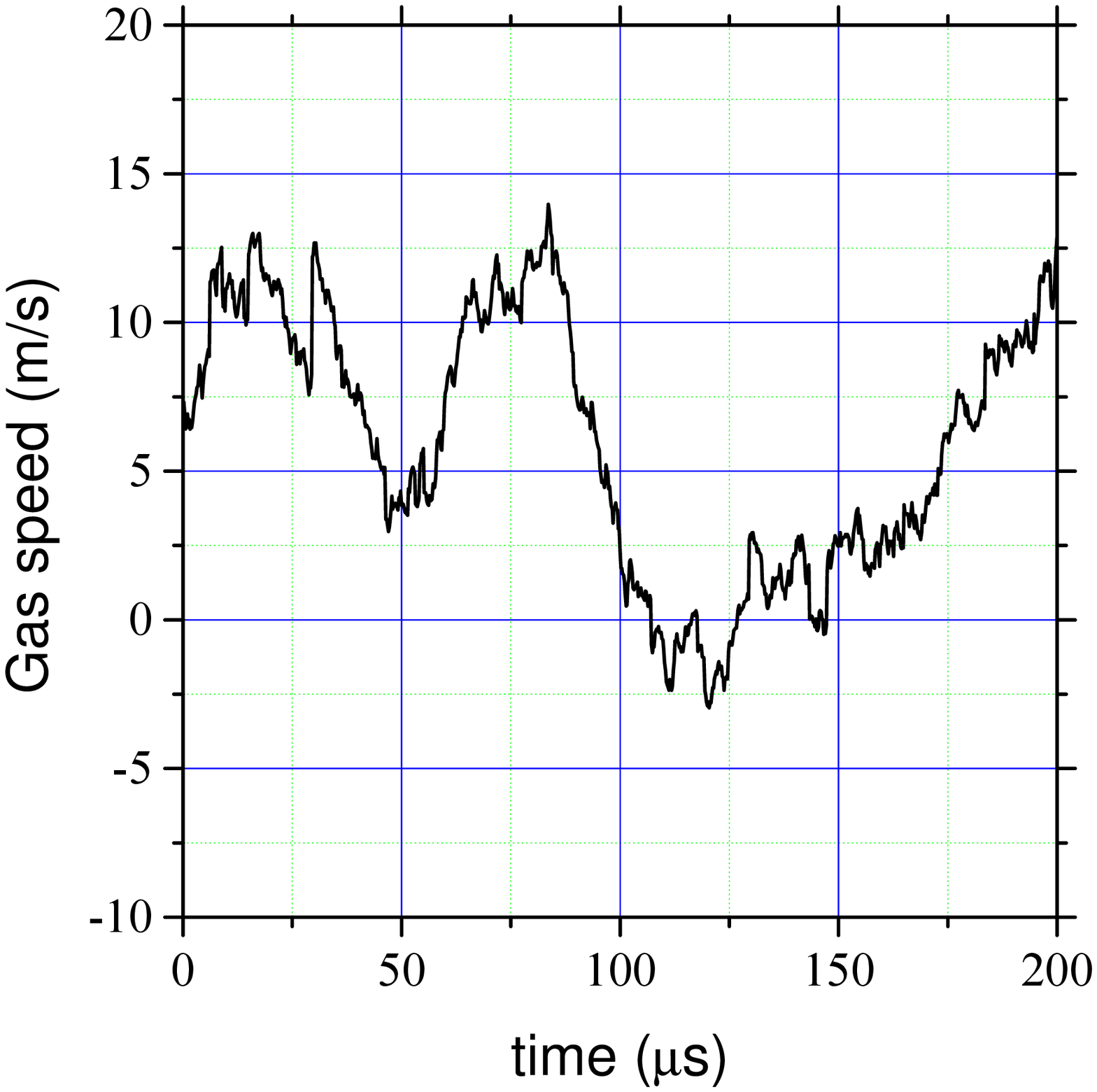}\\
  \caption{Gas speed (in m$/$s) as a function of time.
  Same conditions as in Fig.~\ref{fig4}.}\label{fig5}
\end{figure}

Conditions for maximizing ion-driven gas flow were obtained by Rickard {\it et al.}~\cite{Rickard_06}, and they concluded that, irrespective of geometry, ponderomotive forces on the gas are maximized by increasing current density and by decreasing mobility (i.e., charge carriers which exercise highest drag on the neutral gas). Therefore, $N_2^+$ ions seem to be the best candidate for this purpose.

Our numerical model shows that when the present conditions prevail,
heavy species such as N$_4^+$ move more slowly with the varying external electric field. In fact, the most actives species
on the process of momentum transfer are the electrons and N$_2^+$,
but the molecular ions due to their mass contribute in majority to
control the boundary layer and propulsive force.

From Fig.~\ref{f:small_multiple2} it is noticeable the
multiplications of N$_2^+$ ions when flowing from the electrode edge
to the dielectric surface, flowing along the reverse way as
electrons did. The ions feeding along the dielectric surface is due
to a relative bigger dielectric width which favors the increase of
the ions swarm~\cite{Khudik}, and thus increasing the gas speed due
to the momentum exchange onset from charged particles to neutrals.
Notice that at $t=6 \mu$s nitrogen ions leave the region at the
boundary between the electrode and the dielectric, which corresponds
a region of maximum electric field (see Fig.~\ref{f6}).

Fig.~\ref{fig30} shows contours of constant potential at $t=5 \mu$s.
We can see the decrease of the potential above the energized
electrode and a field reversal region toward the dielectric side
(with the negative electrode below). The negative glow remains in
the proximity of this region, while a second region of field
reversal is also momentously observed. The region of negative
electric potential that appears at 5 $\mu$s in the first half-cycle
is due to the presence of an excess of negative charge due to
O$_2^-$ ions, while the region of maximal electric potential has as
field source all others positive ions. Numerical simulations have
shown that while negative ions in the air do not contribute
significantly to the ponderomotive forces, they can play a role in
the discharge working~\cite{Enloe_06}. Another major structure of
the normal glow discharge remains above with higher values of the
potential. Hence, the phenomenology and typical structures developed
by normal glow discharge are also displayed by the OAUGDP$^{TM}$.
These aspects were also shown in previous publications, like the
one-dimensional numerical simulations of the OAUGDP$^{TM}$ done by
Ben Gadri~\cite{Gadri,Rahel} and fast photography obtained by Massines
{\it et al.}~\cite{Massines}.

Hence, the field reversals observed have the strong contribution of
the negative ions O$_2^-$ which are thus playing an important role
in the discharge working, since it is well known that field reversal
location decides the fraction of ions to the cathode and the
magnitude of the plasma maximum
density~\cite{Boeuf_95,Kolobov,Pinheiro_06,Pinheiro}.

The EHD force acting on the charged particles is given by~\cite{1,6}:
\begin{equation}\label{eq14}
\mathbf{F} = e (n_i - n_e)\mathbf{E} - \nabla (n_i k T_i + n_e k T_e) + (m_e \mathbf{u}_e - m_i \mathbf{u}_i)S,
\end{equation}
where $n_i$ and $n_e$ are respectively the ion and electron number density; $\mathbf{u}_e$ and $\mathbf{u}_i$ are the electron and ion mean velocities; $S$ is the charged particle production rate. The last term of Eq.~\ref{eq14} is here neglected, since its importance confines to other phenomena such as electrophoresis and cathophoresis~\cite{Boeuf_05}, and its contribution to the induced velocity is smaller due to the very small electron$/$neutral atoms mass ratio~\cite{Book1_Roth}. Although we had assumed a constant ion temperature in equilibrium with the gas temperature, we calculated the electron temperature and verified that the order of magnitude of the second term is about 1 $\%$ of the coulombian force term which constitute the main term of the theory of paraelectric gas flow control developed by Roth~\cite{2}. We calculated the electrostriction force term (not included in Eq.~\ref{eq14}) and concluded that is not significant, contributing at maximum with 1 $\%$ to the total ponderomotive force. Subsequently, the ponderomotive forces were averaged over the area of calculation.

It is found that the calculated space averaged
ponderomotive forces per unit volume increases when the electrode
width increases~\cite{Pinheiro_06}. On average, during the second half-cycle the
ponderomotive force magnitude decreases with a magnitude of a few N$/$m as shown in Fig.~\ref{fig4}, a result consistent with experimental results as such presented in Ref.~\cite{Takeuchi}.
This happens when the voltage polarity is reversed and the energized electrode play the
role of cathode. This is due to the potential gradient reduction
on the edge of the expanding plasma (see also Ref.~\cite{10}). This numerical result is contrary to experimental study presented in Ref.~\cite{Hoskinson}, showing that the forces decay exponentially with increasing electrode diameter. This is due to the role of the dielectric that in our model was assumed to absorb electric charged particles instead to feed the swarms and strengthening the body forces (see also Ref.~\cite{Pinheiro_06}). However, there is still no consensus on the ponderomotive force dependency. For example, Singh and Roy~\cite{Singh_08} obtained the magnitude of approximated force and have shown that it increases with the fourth power of the amplitude of the rf potential, implying that the induced fluid velocity also increase. This is certainly an aspect that must be dealt with more caution.

In fact, electrons are faster than ions. After the first
breakdown, they start to charge negatively the dielectric during the 1$st$ half-cycle;
positive ions gains more energy and electrons
are also increasingly accelerated, due to
a higher growing potential. Therefore,
the positive ions density is bigger during
this half-cycle, resulting in stronger
ponderomotive forces.

During the 2$nd$ half-cycle, positive ions (which are mainly formed in volume) tend towards both the electrode and the dielectric
surface; the charge surface density on the dielectric start to become less negative and the gas voltage decrease, generating
less ions and electrons, resulting in smaller body forces; this is the mechanism of an asymmetric flat panel device.

Otherwise, if the discharge is entirely symmetrical in both half-cycles, it is expected that the average gas speed equals zero. In fact, it is the asymmetry in the streamers (in our case bigger during the fist half-cycle) that gives an overall positive gas speed along the axis.

In our present model, calculations of EHD ponderomotive force have shown that its maximum
intensity is attained during electron avalanches, with typical
values on the order of $5 \times 10^9$ N/m$^3$. $F_x$ points along
OX (propelling direction), while $F_y$ points downwards (boundary
layer control). Our calculations show that the resulting average gas speed is about
20 m$/$s and the net EHD body forces (with the present conditions)
have comparable values in the first and second-half cycle, although
slightly bigger during the first half-cycle, as shown in Fig.~\ref{fig5}. It is clear that the
successive streamers that charge the dielectric surface are
responsible for pulling the flow upstream, unidirectionally, as
recent experiments have shown~\cite{Takeuchi}.

\section{Conclusion}

A two-dimensional fluid model of an asymmetric plasma actuator display the behavior of charged species during both half-cycles when electrodes are subject to a sinusoidal applied voltage. The actuator is strongly dominated by N$_2^+$ dynamics, charged species form preferentially at the edge of the electrode with the insulator, and their subsequent behavior and ability to provide an unidirectional gas speed results from the interaction of the charged species with the dielectric, in particular, the effect of the electric field above the insulator and the propensity of the dielectric surface to adsorb or not charged species, and thus controlling the plasma density of the streamers. An appropriate model to describe a realistic interactions of charged species with the dielectric for plasma density enhancement remains to be done, lacking in the literature a more careful study of this important issue.

\begin{acknowledgments}
The authors gratefully acknowledge partial financial support by the
Reitoria da Universidade T\'{e}cnica de Lisboa and the
Funda\c{c}\~{a}o Calouste Gulbenkian. We would also like to thank
important financial support to one of the authors (A.A.M,) in the form of a PhD Scholarship from FCT (Funda\c{c}\~{a}o
para a Ci\^{e}ncia e a Tecnologia).
\end{acknowledgments}

\bibliography{refs}

\begin{thebibliography}{99}

\bibitem{Colver_96} El-Khabiry S. and Colver G. M. 1997 Drag reduction by dc corona discharge along an electrically conductive flat plate for small Reynolds number flow {\it Phys. Fluids} {\bf 9}(3) 587-599

\bibitem{Roth_03} Roth J. R. 2003 Aerodynamic flow acceleration using paraelectric and peristaltic electrohydrodynamic effects of a One Atmosphere Uniform Glow Discharge Plasma {\it Phys. Plasmas} {\bf 10} (5) 2117-2126

\bibitem{Pinheiro_06} Pinheiro M. J. 2006 EHD Ponderomotive Forces and Aerodynamic Flow Control Using Plasma Actuators {\it Plasma Process. Polym.} {\bf 3} 135-141

\bibitem{Moreau_07} Moreau E. 2007 Airflow control by non-thermal plasma actuators {\it J. Phys. D: Appl. Phys.} {\bf 40} (3) 605-636

\bibitem{Soldati_98} Soldati A. and Banerjee S. 1998 Turbulence modification by large-scale organized electrohydrodynamic flows
Alfredo {\it Phys. Fluids} {\bf 10} (7) 1742-1756

\bibitem{Soldati_99} Fulgosi M., Soldati A. and Banerjee S. 1999 Turbulence Modulation by an Array of Large-Scale Streamwise Structures of EHD Origin {\it Proc. of FEDSM'99 3rd ASME/JSME Joint Fluids Engineering Conference} (San Francisco, CA, July 1999) FEDSM99-6934

\bibitem{8} Roth J. R., Madhan R. C. M., Yadav M., Rahel J., and Wilkinson S. P. 2004 Flow Field Measurements of Paraelectric, Peristaltic and Combined Plasma Actuators Based on the One Atmospheric Uniform Glow Discharge Plasma (OAUGDP$^{TM}$) {\it Proc. 42nd Aerospace Sciences Meeting and Exhibit} (Reno, NV) AIAA Paper 2004 - 0845

\bibitem{9} Post M. L. and Corke T. C. 2003 Separation Control on High-Angle-of-Attack Airfoil Using Plasma Actuators AIAA Press (Washington, DC) AIAA Paper 2003-1024

\bibitem{10} Menier E., Leger L., Depussay E., Lago V. and Artana G. 2007 Effect of a dc discharge on the supersonic rarefied air flow over a flat plate {\it J. Phys. D: Appl. Phys.} {\bf 40} 695-701

\bibitem{11} Popovic S. and Vuskovic L. 2006 Aerodynamic effects in weakly ionized gas: phenomenology and applications in {\it The Physics of Ionized Gases} AIP Conference Proceeindings {\bf 876} 272-283

\bibitem{Sigelman_70} David Sigelman 1970 Sonic boom minimization schemes {\it J. Aircraft} {\bf 7} (3) 280

\bibitem{Boyd_99} Cain T., Boyd D. 1999 Electroaerodynamics and the Effect of an Electric Discharge on Cone/Cylinder Drag at Mach 5  {\it 37th Aerospace Sciences Meeting and Exhibit} (Reno, NV) AIAA-99-0602

\bibitem{12} Huang X., Chan S. and Zhang X. 2007 Atmospheric plasma actuators for aeroacoustic applications {\it IEEE Trans. Plasma Sci.} {\bf 35} (3) 693-695

\bibitem{13} Chan S., Zhang X. and Gabriel S. 2005 The Attenuation of Cavity Tones Using Plasma Actuators 11th AIAA/CEAS Aeroacoustics Conference {\it 26th AIAA Aeroacoustics Conference} (Monterey, CA) AIAA Paper 2005-2802

\bibitem{14} J. R. Roth 1999 ``Method and apparatus for cleaning surfaces with a glow discharge plasma at one atmosphere of pressure", US Patent 5,938,854

\bibitem{Shin_07} Shin J., Narayanaswamy V., Laxminarayan V., Raja L. and Clemens N. T. 2007 Characterization of a Direct-Current Glow Discharge Plasma Actuator in Low-Pressure Supersonic Flow  {\it AIAA Journal} {\bf 45} (7) 1596-1605

\bibitem{Utkin_07} Utkin Y. G., Keshav S., Kim J.-H., Kastner J., Adamovich I. G. and Samimy Mo 2007 Development and use of localized arc filament plasma actuators for high-speed flow control {\it J. Phys. D: Appl. Phys.} {\bf 40} 685-694

\bibitem{1} Font G. I. 2004 Boundary Layer Control with Atmospheric Plasma Discharge {\it 40th AIAA/ASME/SAE/ASEE Joint Propulsion Conference and Exhibit} (Fort Lauderdale, Fl) AIAA Paper 2004-3574

\bibitem{2} Font G. I. and Morgan W. L. 2005 Plasma Discharges in Atmospheric Pressure Oxygen for Boundary Layer Separation Control
           {\it 35th AIAA Fluid Dynamics Conference and Exhibit} (Toronto, Ontario) AIAA Paper 2005-4632

\bibitem{4} Font G. I., Jung S., Enloe C. L., McLaughlin T. E., Morgan W. L. and Baughn J. W. 2006 Simulation of the effects of force and heat produced by a plasma actuator on neutral flow evolution {\it 44th AIAA Aerospace Sciences Meeting and Exhibit} (Reno, NV) AIAA Paper 2006-0167

\bibitem{5} J. P. Boeuf, Y. Lagmich, Th. Unfer, Th. Callegari, and L. C. Pitchford 2007 Electrohydrodynamic force in dielectric barrier discharge plasma actuator {\it J. Phys. D: Appl. Phys.} {\bf 40} (3) 652-662

\bibitem{6} J. P. Boeuf, and L. C. Pitchford 2005 Electrohydrodynamic force and aerodynamic flow acceleration in surface dielectric barrier discharge {\it J. Appl. Phys.} {\bf 97} 103307

\bibitem{15} G\"{o}ksel B., Greenblatt D., Rechenberg I., Singh Y., Nayeri C. N. and Paschereit C. O. 2006 Pulsed plasma actuators for separation flow control {\it Conference on Turbulence and Interactions TI2006} (Porquerolles, France)

\bibitem{Enloe_03} Enloe C L, McLaughlin T E, VanDyken R D, Kachner K D, Jumper E J and
Corke T C 2003 Mechanisms and responses of a single dielectric barrier plasma {\it AIAA}
2003-1021

\bibitem{Enloe_06} Font G I, Jung S, Enloe C L, McLaughlin T E, Morgan W L and Baughn J W
2006 Simulation of the Effects of Force and Heat Produced by a Plasma Actuator on Neutral
Flow Evolution {\it AIAA} 2006-0167

\bibitem{Singh} Singh K P, Subrata R and Gaitonde D V 2006 Modeling of Dielectric Barrier
Discharge Plasma Actuator with Atmospheric Air Chemistry {\it AIAA} 2006-3381

\bibitem{Gadri} Gadri R B 1999 One atmosphere glow discharge structure revealed by
computer modeling{\it IEEE Trans. Plasma Sci.} {\bf 27} 36-37
\bibitem{Kossyi92} Kossyi I A, Kostinsky A Y, Matveyev A A, and Silakov V P 1992 Kinetic
scheme of the non-equilibrium discharge in nitrogen-oxygen mixtures {\it Plasma Sources
Sci. Technol.} {\bf 1} 207-220

\bibitem{Ferreira} Ferreira C M, Alves L L, Pinheiro M and S\'{a} P A 1991 Modeling of
Low-Pressure Microwave Discharges in Ar, He and O2: Similarity Laws for the maintenance
field and mean powertransfer{\it IEEE Trans. Plasma Sci.} {\bf 19} 229-239

\bibitem{Siglo} Siglo Data Base: http:cpat.ups-tlse.fr

\bibitem{Sigmond} Sigmond R S 1979 Gas Discharge Data Sets for Dry Oxygen
and Air {\it Electron and Ion Physics Research Group Report} The
Norwegian Institute of Technology, The University of Trondheim

\bibitem{Massines_99} Gouda G and Massines F 1999 Role of excited species in dielectric
barrier discharge mechanisms observed in helium at atmospheric pressure{\it Electrical
Insulation and Dielectric Phenomena Annual Report Conference} {\bf 2} 496-499

\bibitem{Patankar} Patankar S V 1980 {\it Numerical heat transfer and fluid
flow}, (New York: Taylor $\&$ Francis)

\bibitem{Sato} Morrow R and Sato N 1999 The discharge current induced by the motion of
charged particles in time-dependent electric fields; Sato's equation extended  {\it J.
Phys.} D: {\it Appl. Phys.} {\bf 32} L20-L22

\bibitem{Gibalov_04} Gibalov V I and Pietsch G J 2004 Properties of dielectric barrier
discharges in extended coplanar electrode systems {\it J. Phys.} D: {\it Appl. Phys.}
{\bf 37} (15) 2093-2100

\bibitem{Rickard_06} Rickard M, Dunn-Rankin D, Weinberg F and Carleton F 2006 Maximizing
ion-driven gas flows {\it J. Electrostat.} {\bf 64} (6) 368-376

\bibitem{Khudik} Shvydky A, Nagorny V P and Khudik V N 2004 The electron avalanche
sliding along the dielectric surface{\it J. Phys.} D: {\it Appl. Phys.} {\bf 37} 2996-2999

\bibitem{Rahel} R\'{a}hel J and Sherman D M 2005 The transition from a filamentary
dielectric barrier discharge to a diffuse barrier discharge in air at atmospheric
pressure {\it J. Phys.} D: {\it Appl. Phys.} {\bf 38} (4) 547-554

\bibitem{Massines} Massines F, Rabehi A, Decomps Ph, Gadri R B, Segur P and Mayoux Ch
1998 Mechanisms of a Glow Discharge at Atmospheric Pressure Controlled by Dielectric
Barrier{\it J. Appl. Phys.} {\bf 83} (6) 2950-2957

\bibitem{Boeuf_95} Boeuf J P and Pitchford L C 1995 Field reversal in the negative glow
of a dc glow discharge{\it J. Phys.} D: {\it Appl. Phys.}{\bf 28} 2083-2088

\bibitem{Kolobov} Kolobov V I and Tsendin L D 1992 Analytic model of the cathode region
of a short glow discharge in light gases {\it Phys. Rev.} A {\bf 46} (12) 7837-7852

\bibitem{Pinheiro} Pinheiro M J 2004 {\it Phys. Rev.} E {\bf 70} 056409; Pinheiro M J
2007 {\it Gas Discharges-Fundamentals and applications}, ed. Jayr de Amorim Filho,
Transworld Research Network, Kerala, India

\bibitem{Boeuf_05} Boeuf J P and Pitchford L C 2005 Electrohydrodynamic force and
aerodynamic flow acceleration in surface dielectric barrier discharges{\it J. Appl.
Phys.} {\bf 97} (10) 103307-103307-10

\bibitem{Book1_Roth} Roth J R 2001 {\it Industrial Plasma Engineering}, {\bf 2}, 235
(IOP, Bristol)

\bibitem{Takeuchi} Takeuchi N, Yasuoka K and Ishii S 2007 Measurement of instantaneous
Gas Flow Induced by Plasma Actuator with Laser Doppler Velocimeter {\it Proc. 18th
International Symposium on Plasma Chemistry} {\bf 29C-a5} 645 Kyoto, Ed. K. Tachibana, O.
Takai, K. Ono, T. Shirafuji

\bibitem{Hoskinson} Hoskinson A R, Hershkowitz N and Ashpis D E 2008 Force measurements
of single and double barrier DBD plasma actuators in quiescent air
 {\it J. Phys.}D: {\it Appl. Phys.} {\bf 41} (24) 245209-245209-9

\bibitem{Singh_08} Singh K P and Subrata R 2008 Force approximation for a plasma actuator
operating in atmospheric air{\it J. Appl. Phys.} {\bf 103} 013305-013305-6

\end{thebibliography}
\end{document}